\documentclass{aa}  

\usepackage{graphicx}
\usepackage{float,dblfloatfix}   
\usepackage[varg]{txfonts}
\usepackage{natbib,twoopt}
\usepackage[breaklinks=true,draft]{hyperref} 
\bibpunct{(}{)}{;}{a}{}{,}             
\makeatletter
  \newcommandtwoopt{\citeads}[3][][]{\href{http://adsabs.harvard.edu/abs/#3}%
    {\def\hyper@linkstart##1##2{}%
     \let\hyper@linkend\@empty\citealp[#1][#2]{#3}}}
  \newcommandtwoopt{\citepads}[3][][]{\href{http://adsabs.harvard.edu/abs/#3}%
    {\def\hyper@linkstart##1##2{}%
     \let\hyper@linkend\@empty\citep[#1][#2]{#3}}}
  \newcommandtwoopt{\citetads}[3][][]{\href{http://adsabs.harvard.edu/abs/#3}%
    {\def\hyper@linkstart##1##2{}%
     \let\hyper@linkend\@empty\citet[#1][#2]{#3}}}
  \newcommandtwoopt{\citeyearads}[3][][]%
    {\href{http://adsabs.harvard.edu/abs/#3}
    {\def\hyper@linkstart##1##2{}%
     \let\hyper@linkend\@empty\citeyear[#1][#2]{#3}}}
\makeatother
\begin{document}

   \title{A non-uniform distribution of the nearest brown dwarfs}

   \titlerunning{A non-uniform distribution of the nearest brown dwarfs}

   \author{G.~Bihain\inst{1}
          \and R.-D. Scholz\inst{1}
           }

   \authorrunning{Bihain}
   \institute{Leibniz-Institut f{\"u}r Astrophysik Potsdam (AIP), An der Sternwarte 16, 
              14482 Potsdam, Germany\\
            \email{[gbihain,rdscholz]@aip.de}
             }

   \date{Received 19 December 2015 / Accepted 23 February 2016}

 
  \abstract
   {The census of solar neighbours is still complemented by new discoveries, mainly of very low-mass,
faint dwarfs, close to or within the substellar domain. These discoveries contribute to a better understanding
of the field population; its origin in terms of Galactic dynamics and (sub)stellar formation and evolution.
Also, the nearest stars and brown dwarfs at any given age allow the most precise direct characterization,
including the search for planetary companions.}
   {We aim to further assess the substellar census on the Galactic plane.}
   {We projected the 136 stars and 26 brown dwarfs known at $<$6.5~pc on the Galactic plane and evaluated their distributions.}
   {Stars present a uniform- and brown dwarfs a non-uniform distribution, with 21 objects behind the Sun and only five
ahead relative to the direction of rotation of the Galaxy. This substellar configuration has a
probability of 0.098$^{+10.878}_{-0.098}$\% relative to uniformity. The helio- and geocentric nature of the distribution suggests it might
result in part from an observational bias, which if compensated for by future discoveries, might increase the
brown-dwarf-to-star ratio, shifting it closer to values found in some star forming regions.}
  {}


   \maketitle
%

\section{Introduction}

The stellar neighbourhood of the Sun comprises main sequence stars in the majority and a minority of cool
white dwarfs (see RECONS\footnote{REsearch Consortium On Nearby Stars, \url{www.recons.org}} census)
spanning ages of $\sim$1-10~Gyr \citepads{2012ApJS..199...29G}. At $<$6.5~pc, the stellar census is
expected to be relatively complete with a few new white- or low-mass dwarfs to be discovered. The
first substellar objects identified at $<$6.5~pc were \object{GJ~229~B}, a methane dwarf (T7V;
\citeads{1995Natur.378..463N,1995Sci...270.1478O}), and \object{LP~944-020}, a lithium brown dwarf
(M9; \citeads{1998MNRAS.296L..42T}). Searches based on the Two Micron All Sky Survey (2MASS,
\citeads{2006AJ....131.1163S}) and Wide-field Infrared Survey Explorer (WISE,
\citeads{2010AJ....140.1868W}) as well as other surveys (see review by
\citeads{2012ApJ...753..156K}) have incremented the number of brown dwarfs. Star forming
regions appear to produce free-floating substellar objects in abundance, even at planetary masses
\citepads{2001ApJ...556..830B,2012ApJ...754...30P,2016A&A...586A.157P,2015ApJ...810..159M};
furthermore, the lower their masses, the higher their velocity dispersions
\citepads{2014A&A...568A..77Z}. This explains why numerous faint and (infra)red relics of such
objects are indeed found in the solar vicinity in the form of ultra-cool, late-T- and early-Y-type
dwarfs, with the $\sim$3--10~$M_{\rm jup}$ \object{WISE~0855-0714} as the most spectacular example
\citepads{2014ApJ...786L..18L}. To further assess the substellar census, we consider the
distribution of known objects on the Galactic plane.

\begin{figure*}[th!] \resizebox{\hsize}{!}{\includegraphics{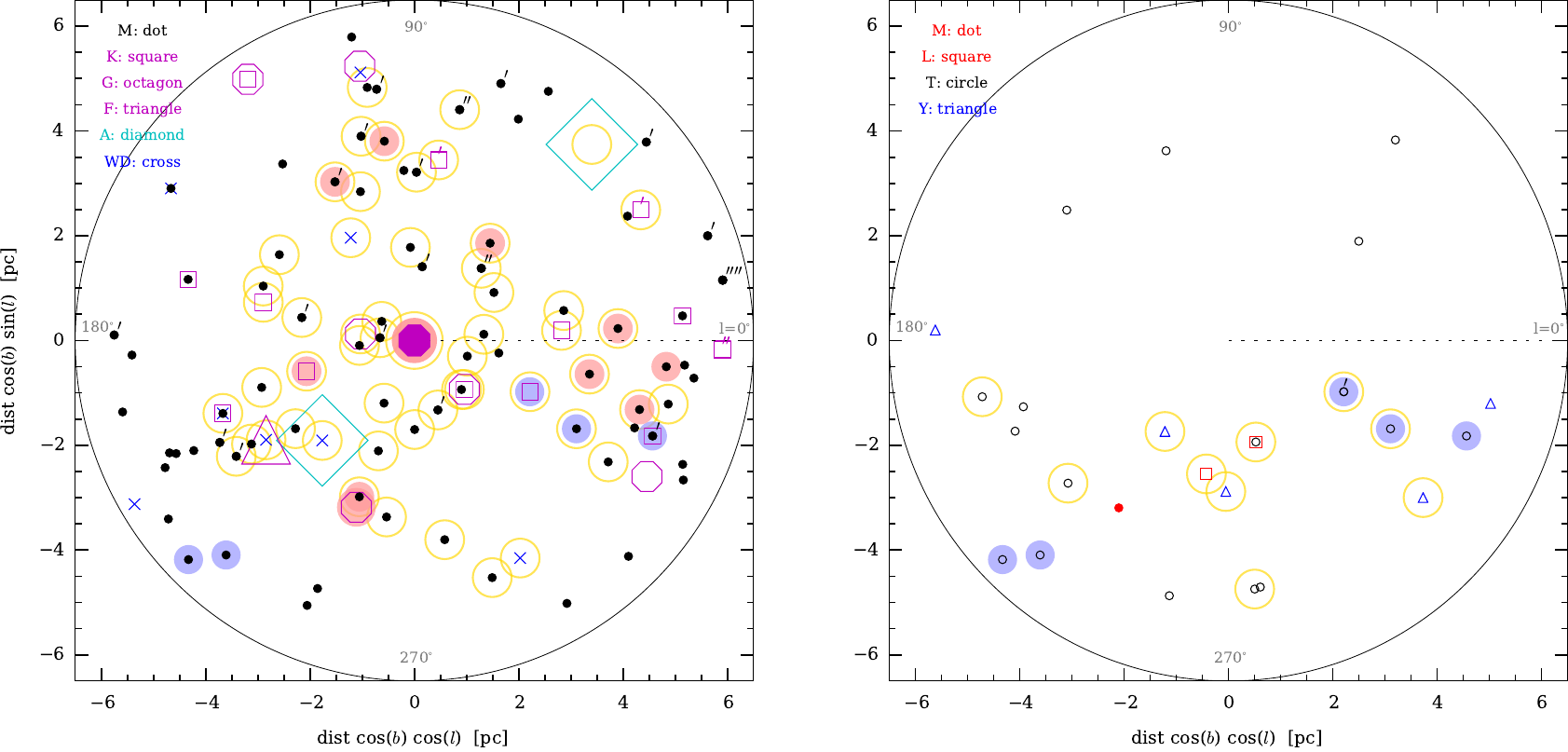}} \caption{Distribution of
$<$6.5\,pc  stars (left panel) and brown dwarfs (right panel) on the Galactic coordinates plane. M, K, G,
F, A, and WD type stars are represented by dots, squares, octagons, triangles, diamonds, and crosses, and
M, L, T, and Y type brown dwarfs are represented by dots, squares, circles, and triangles, respectively.
The filled octagon indicates the Sun and the dotted line the direction towards the Galactic centre.
Different object symbols centred on the same location are multiple systems; a prime indicates one more
component of the same spectral type. Symbols with blue and pink surroundings are star-brown dwarf and
star-planet systems, respectively. Single objects or multiple systems nearer than 5.16~pc (half-volume
distance) are indicated with a yellow circle.} \label{distr} \end{figure*}


\section{Data analysis and results}\label{results}

We updated the RECONS list (as of 2012 January 1) of the 100 nearest systems ($<$6.5~pc) and the
8~pc sample by \citetads{2012ApJ...753..156K} with new objects or improved parallaxes or spectral
types. At $<$6.5~pc, 16 brown dwarfs (Table~\ref{bdtab}) are not in RECONS\\

\begin{table}[H]
\tiny
\protect\caption[]{Brown dwarfs known at $<$6.5~pc.}
\label{bdtab}
\centering          
     \tabcolsep=0.187cm
     \begin{tabular}{l c l l l}
         \hline\hline
         Object $^a$ &
         \,$l$~~~~~~~~~~~~$b$ &
         Plx &
         SpT &
         Ref$^b$
         \\
         &\,(deg)~~~~~(deg)& (mas) & & 
         \\
         \hline
\object{WISE J1049-5319 A}  &  285.18~     5.29 & 496\,$\pm$\,37             & L7.5 & 1,2   \\  
\object{WISE J1049-5319 B}  &  285.18~     5.29 & 496\,$\pm$\,37             & T0.5 & 1,2   \\  
\object{WISE J0855-0714  }  &  234.99~    23.35 & 433\,$\pm$\,15             & $>$Y2& 3,4   \\  
\object{eps Indi Ba      } *&  336.12~~  -48.15 & 276.06\,$\pm$\,0.28        & T1   & 5,6   \\  
\object{eps Indi Bb      } *&  336.12~~  -48.15 & 276.06\,$\pm$\,0.28        & T6   & 5,6   \\  
\object{SCR J1845-6357 B } *&  331.52~~  -23.50 & 259.50\,$\pm$\,1.11        & T6   & 7,8   \\  
\object{UGPS J0722-0540  }  &  221.54~     4.27 & 242.80\,$\pm$\,24          & T9   & 9,10  \\  
\object{WISE J0350-5658  }  &  269.01~~  -46.67 & 238\,$\pm$\,38             & Y1   & 11,11 \\  
\object{DENIS J0817-6155 }  &  276.06~~  -14.40 & 203\,$\pm$\,13             & T6   & 12,12 \\  
\object{WISE J1639-6847  }  &  321.20~~  -14.48 & 202.3\,$\pm$\,3.5          & Y0   & 13,14 \\  
\object{DENIS J0255-4700 } *&  260.59~~  -58.67 & 201.37\,$\pm$\,3.89        & L9   & 15,16 \\  
\object{WISE J0521+1025  }  &  192.82~~  -14.62 & 200\,$\pm$\,52             & T7.5 & 17,17 \\  
\object{WISE J1506+7027  }  &  108.27~    42.61 & 193\,$\pm$\,26             & T6   & 11,18 \\  
\object{2MASS J0939-2448 } *&  256.91~    20.47 & 187.30\,$\pm$\,4.60        & T8   & 19,19 \\  
\object{WISE J1741+2553  }  &  50.11~     26.10 & 180\,$\pm$\,15             & T9   & 20,21 \\  
\object{2MASS J1114-2618 }  &  277.39~    31.73 & 179.20\,$\pm$\,1.4         & T7.5 & 22,16 \\  
\object{2MASS J0415-0935 } *&  202.94~~  -38.94 & 175.20\,$\pm$\,1.70        & T8   & 22,16 \\  
\object{WISE J1541-2250  }  &  346.55~    25.26 & 175.1\,$\pm$\,4.4          & Y0.5 & 13,11 \\  
\object{GJ 229 B         } *&  228.61~~  -18.44 & 173.81\,$\pm$\,0.99        & T7   & 5,16  \\  
\object{GJ 570 D         } *&  338.23~    32.75 & 171.22\,$\pm$\,0.94        & T7.5 & 5,16  \\  
\object{WISE J0720-0846 B}  &  224.01~     2.29 & 166\,$\pm$\,28             & T5.5 & 23,24 \\  
\object{2MASS J0937+2931 } *&  197.85~    47.58 & 163.39\,$\pm$\,1.76        & T6   & 25,16 \\  
\object{WISE J0410+1502  }  &  177.95~~  -25.92 & 160\,$\pm$\,9              & Y0   & 26,27 \\  
\object{2MASS J1503+2525 }  &  37.19~     60.47 & 157.2\,$\pm$\,2.2          & T5   & 22,16 \\  
\object{SIMP J0136+0933  }  &  141.22~~  -51.69 & 156\,$\pm$\,7              & T2.5 & 28,28 \\  
\object{LP 944-020       } *&  236.67~~  -53.41 & 155.89\,$\pm$\,1.03        & M9   & 29,30 \\  
        \hline
        \hline
      \end{tabular}
\begin{flushleft}
$^a$ *: In RECONS.
$^b$ References for parallaxes and spectral types are:
(1) \citetads{2013ApJ...767L...1L}
(2) \citetads{2013ApJ...772..129B}
(3) \citetads{2014ApJ...796....6L}
(4) \citetads{2015ApJ...799...37L}
(5) \citetads{2007A&A...474..653V}
(6) \citetads{2004A&A...413.1029M}
(7) \citetads{2006AJ....132.2360H}
(8) \citetads{2007A&A...471..655K}
(9) \citetads{2012ApJ...748...74L}
(10) \citetads{2010MNRAS.408L..56L}
(11) \citetads{2012ApJ...753..156K}
(12) \citetads{2010ApJ...718L..38A}
(13) \citetads{2014ApJ...796...39T}
(14) \citetads{2015ApJ...804...92S}
(15) \citetads{2006AJ....132.1234C}
(16) \citetads{2006ApJ...637.1067B}
(17) \citetads{2013A&A...557A..43B}
(18) \citetads{2011ApJS..197...19K}
(19) \citetads{2008ApJ...689L..53B}
(20) \citetads{2013Sci...341.1492D}
(21) \citetads{2011A&A...532L...5S}
(22) \citetads{2012ApJS..201...19D}
(23) \citetads{2015AJ....149..104B}
(24) \citetads{2015AJ....150..180B}
(25) \citetads{2009A&A...493L..27S}
(26) \citetads{2014ApJ...783...68B}
(27) \citetads{2011ApJ...743...50C}
(28) \citetads{2006ApJ...651L..57A}
(29) \citetads{2014AJ....147...94D}
(30) \citetads{2004AJ....128.2460H}
\end{flushleft}
\end{table}

\noindent yet. (The RECONS census is restricted to objects with trigonometric parallax errors smaller than
10~mas.) Similarly, eight stars at $<$6.5~pc are not in RECONS. \object{WISE~J1540-5101} (M7;
\citeads{2014ApJ...783..122K,2014A&A...567A...6P}), \object{TYC 3980-1081-1} ($\sim$M3.5;
\citeads{2014AJ....148..119F}), and \object{WISE~J0720-0846 AB} (M9.5+T5;
\citeads{2014A&A...561A.113S,2015A&A...574A..64I,2015AJ....149..104B,2015AJ....150..180B}) are three
recently discovered stars. Both \object{2MASS J0533-4257} (M4.5) and \object{2MASS J1845-1409 AB} (M5+M5)
have provisional distances \citepads{2012ApJ...753..156K}. Finally, \object{G\,161-71} and
\object{L\,43-72} (M6 and M5 as from \citeads{2014MNRAS.443.2561G}), already mentioned as nearby stars by
\citetads{2002A&A...390L..27M} and \citetads{2005A&A...439.1127S}, respectively, have updated distances
\citepads{2015AJ....149....5W}. Compared to the sample of \citetads{2012ApJ...753..156K},  at $<$6.5~pc
there are six new brown dwarfs (\object{WISE~J1049-5319 AB}, \object{WISE~J0855-0714},
\object{WISE~J1639-6847}, \object{WISE~J0521+1025}, \object{WISE~J0720-0846 B}), and owing to more
accurate parallaxes, one brown dwarf shifting in (\object{2MASS~J1114-2618}), three shifting out but at
$<$8~pc (\object{WISE~J2056+1459}, \object{WISE~J1405+5534}, and \object{WISE~J0254+0223}; see references
below), and two shifting out at $>$8~pc (\object{WISE~J0146+4234} and \object{WISE~J0535-7500};
trigonometric parallaxes from \citeads{2014ApJ...783...68B} and \citeads{2014ApJ...796...39T},
respectively).

We adopted the stellar- or brown dwarf status of the objects as given in the literature. The broad range
of age estimates for the nearby brown dwarfs (a few 100~Myr to about 10 Gyr) and the substellar
mass-age-luminosity degeneracy imply that masses cannot be obtained accurately, except for short period
binaries, for which dynamical masses can be measured. To simplify, we considered i) a solar metallicity,
stellar-brown dwarf boundary mass (0.075$M_{\odot}$, \citeads{2000ARA&A..38..337C}); (ii) all substellar
objects of Table~\ref{bdtab} as part of the brown dwarf group, even if some might reach planetary-masses,
below the theoretical minimum mass for deuterium burning ($\sim$0.013$M_{\odot}$,
\citeads{1996ApJ...460..993S}); and (iii) stars and brown dwarfs as two groups, without studying their
mass overlap due to age, metallicity, or dynamical mass uncertainties further. Nevertheless, we recall
that low-mass stars are predicted to reach effective temperatures $\ga$2000~K (as estimated using the
10~Gyr Dusty model from \citeads{2000ApJ...542..464C}, and corresponding to optical spectral types
$\la$L3, using the scale from \citeads{2008ApJ...689.1295K}) and that the cooler, late-type L, T, and Y
dwarfs of the sample are therefore expected to be substellar objects, independent of age (see also
\citeads{2013AN....334...26K}). The L7.5 and T0.5 components of \object{WISE~J1049-5319} in the 6.5~pc
sample, for instance, present \ion{Li}{I} absorption in their optical spectra and are brown dwarfs
\citepads{2014ApJ...790...90F,2015A&A...581A..73L} based on the lithium test
\citepads{1992ApJ...389L..83R}. Besides, LP~944-20 (M9) is the unique M-type object in the 6.5~pc sample
identified as a brown dwarf, also through the lithium test. Accounting for the most massive star,
\object{Sirius~A} (A1.0~V; 2.03$M_{\odot}$, \citeads{2013MNRAS.435.2077H}), the stellar mass range is
2--0.075$M_{\odot}$; we note that the stellar dwarf with the latest spectral type,
\object{WISE~J0720-0846~A} (M9.5~V), has a dynamical mass estimate of 0.08$M_{\odot}$
\citepads{2015AJ....150..180B}. Accounting for the least massive substellar object,
\object{WISE~0855-0714} ($>$Y2, \citeads{2014ApJ...786L..18L}), the substellar mass range is
0.075--0.007$M_{\odot}$.


 
We represent the 136 stars (including the Sun) and 26 brown dwarfs on the Galactic plane, in the left-
and right-hand panels of Fig.~\ref{distr}, respectively. These stars distribute uniformly with about 50\%
ahead ($l$=0--180~deg) and 50\% behind ($l$=180--360~deg) the Sun relative to the direction of Galaxy
rotation ($l$=90~deg). On the other hand, brown dwarfs do not distribute uniformly: there are 21
(81\%) behind and only five (19\%) ahead of the Sun (including the Y dwarf at $l$=178~deg). Both the T-
and Y-type brown dwarfs present a deficit at $l$=0--180~deg: four of the five Y dwarfs are at
180--360~deg, while there are 14 T dwarfs at 180--360~deg compared to only four T dwarfs at
$l$=0--180~deg. Even the two late-L- and the one late-M-type brown dwarfs are at $l$=180--360~deg
(near $l$$\sim$270~deg). Besides this, the six (T-type) brown dwarfs that are companions to five stars
(symbols with blue surroundings in Fig.~\ref{distr}) are also at $l$=180--360~deg.

The same stars and brown dwarfs projected on the orthogonal plane to the Galactic plane and along $l=0$~deg
(see Fig.~A.1) distribute relatively uniformly. As in the previous projection, because the initial
volume is a sphere, the space sampled transversally is smaller at larger radii, in contrast to a
cylindrical volume, which explains the fewer objects. Among the nearest objects, 48.5\% (66/136) of the
stars belong to stellar multiple systems, while 15.4\% (4/26) of the brown dwarfs belong to brown dwarf
multiple systems (see Figs.~\ref{distr} and A.1). At least five stars have brown dwarf companions
and 10 stars other than the Sun have confirmed planets (based on the Open Exoplanet
Catalogue\footnote{{\url{www.openexoplanetcatalogue.com}}, as on 2016 February 9.}), implying percentages
of $>$3.7 and $>$8.1\%, respectively. The number of substellar companions found to the stars and brown
dwarfs will probably increase with the sensitivity of the low-mass companion searches. 

Assuming a uniform distribution of brown dwarfs on the Galactic plane, as observed for stars and as also
observed on the perpendicular plane for both brown dwarfs and stars, we would expect a symmetry between the
two halves of the projected sphere, split by the galactocentric line crossing the Sun. The probability of
the observed configuration can then be estimated using the binomial distribution $P(k; n, p)$ =
$n!/(k!(n-k)!)$ * $p^k$ * $(1-p)^{n-k}$ = $n!/(k!(n-k)!)$ * $0.5^n$, where $k$ is the number of brown
dwarfs ahead of the Sun, $n$  the total number of brown dwarfs, and $p=0.5$  the probability of an object
being ahead of the Sun for the assumed symmetric distribution. Thus $P(5; 26)=0.09802$\%. Excluding the six
brown dwarf companions to stars, the configuration probability for the substellar population unbound to
stars would be $P(5; 20)=1.47858$\%.


Subtracting or adding the uncertainties to the parallaxes implies larger or smaller distances and thus may
decrease or increase the number of objects within 6.5~pc, respectively. Because some parallaxes may present
larger errors than those quoted (e.g. for preliminary trigonometric parallaxes based on few epochs or
objects that are multiple and thus farther than the assumed single-object spectrophotometric distance), we
considered subtracting or adding twice the quoted uncertainty. In the first case, the space ahead of the
Sun is left void ($k$=0), the front Y0-type \object{WISE~J0410+1502}, and the four T-type
\object{WISE~J1506+7027}, \object{WISE~J1741+2553}, \object{2MASS~J1503+2525}, and \object{SIMP~J0136+0933}
having shifted out of the $<$6.5~pc volume, while $n$=18 brown dwarfs remain behind the Sun (two T- and the
one M-type dwarf shifting out, too). In the second case, 12 objects shift in ($n$=38), nine ahead (a), and
three behind (b) the Sun ($k$=14), namely \object{WISE~J0313+7807}, \object{WISE~J2056+1459}, and
\object{WISE~J2209+2711} (T8.5, Y0, and Y1, \mbox{--a--;} trigonometric parallaxes from
\citeads{2014ApJ...783...68B}), \object{WISE~J1405+5534} and \object{WISE~J0254+0223} (Y0 and T8,
\mbox{--a--;} trigonometric parallaxes from \citeads{2013Sci...341.1492D}), \object{WISE~J1928+2356},
\object{WISE~J0005+3737}, \object{WISE~J0049+2151}, \object{WISE~J0607+2429}, and \object{2MASS~J0348-6022}
(T6, T9, T8.5, L9, and T7, \mbox{--aaabb--;} spectro-photometric distances from
\citeads{2012ApJ...753..156K}, where we assume 1-$\sigma$ errors of 1~pc), and \object{WISE~J2030+0749} and
\object{WISE~J0457-0207} (T1.5 and T2, \mbox{--ab--;} spectro-photometric distances from
\citeads{2013A&A...557A..43B}). Assuming that the distance estimates of the objects are independent, we can
compare the minimum number of objects ahead ($k$=0) to the maximum number of objects behind (24) to obtain
the lower limit probability, $P(0; 24)=0.000006$\%. Conversely, we can compare the maximum number of
objects ahead ($k$=14) to the minimum number of objects behind (18) to obtain the higher limit probability,
$P(14; 32)=10.97647$\%.

Therefore, the observed brown dwarf configuration would have a probability with 2-$\sigma$ error bars
of 0.098$^{+10.878}_{-0.098}$\%.

Finally, the five brown dwarfs at $l$=0--180~deg are farther at 5.2--6.5~pc, implying the brown dwarf
distribution at $\la$5~pc is even less uniform (as highligted in Figs.~\ref{distr} and A.1).
Indeed, at less than the half-volume distance of 5.16~pc, the brown dwarf numbers reduce to zero and 12
ahead and behind the Sun, respectively, compared to five and nine at 5.16--6.5~pc in the outer shell
volume, while the stellar distribution remains relatively uniform. Furthermore, at $<$5.16~pc, the brown
dwarfs are close to or below the Galactic height of the Sun. Extending the 6.5~pc sample to a less complete
and less accurate 8~pc sample by including farther brown dwarfs (as above), we find eight and three
additions ahead and behind the Sun, and thus in total 13 and 24 brown dwarfs, respectively, indicating
about half less substellar objects ahead of the Sun ($P(13; 37)=2.592$\%). Using the initial 8~pc data of
\citetads{2012ApJ...753..156K}, we similarly find 12 and 21 brown dwarfs.


\section{Discussion}

\begin{figure}[h!] \resizebox{\hsize}{!}{\includegraphics{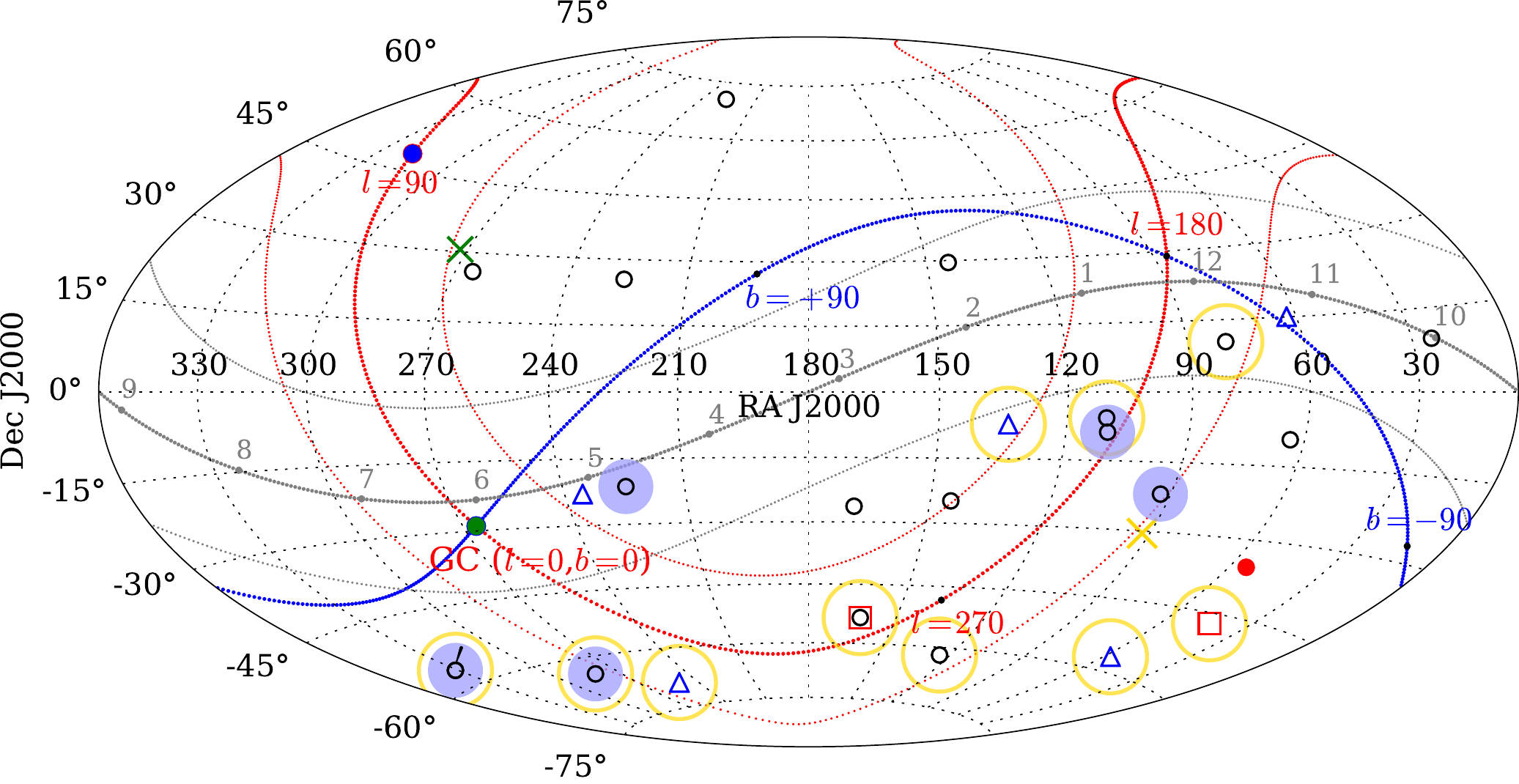}} \caption{Equatorial coordinate sky
map in the Hammer equal-area projection of brown dwarfs at $<$6.5\,pc. Same symbols as in
Fig.~\ref{distr}. The Galactic plane (with $b=\pm20$~deg parallels) and its orthogonal plane along
$l=0$~deg are represented by the red and blue lines, while the Galactic centre and the direction of
Galactic rotation are indicated by the green and blue filled circles, respectively. The region that is
particularly void of brown dwarfs is above the blue line ($l\sim0-180$~deg). The solar apex and antapex
are indicated by green and yellow crosses. The path of the Sun (ecliptic) with $\beta$=$\pm$20~deg
parallels are represented by the grey lines. The month (approximate mid-month date) of optimal night
visibility from observatories such as at Mauna Kea or Cerro Paranal for each emphasized sky location
on the ecliptic are indicated.} \label{skymap} \end{figure}

To discuss the brown dwarf distribution further, we represent these objects in equatorial
coordinates on a Hammer equal-area projection, as shown in Fig.~\ref{skymap}. Here also we see a
clear void that is well defined above the blue line (orthogonal plane to the Galactic plane, along
$l=0$~deg). It coincides in part with the northern hemisphere. Furthermore, there are no brown
dwarfs found yet within $-20<b<20$~deg of the Galactic plane (between the two outer red lines) and
at $l=0-180$~deg.

We wonder whether the non-uniform distribution of the brown dwarfs at $<$6.5~pc could be related to (i) a
small number bias, (ii) incompleteness of all-sky surveys, (iii) incompleteness of the searches, and (iv)
"brown dwarf" statistics or Galactic dynamics:

(i) While the number of brown dwarfs is small when divided into spectral classes (1M, 2L, 18T, and 5Y), the
total number (26) is large and representative enough for statistical estimations, and it indicates that the
distribution is remarkably non-uniform. We note that the 17 K-type dwarfs, almost equal in number to the 18
T dwarfs, do not distribute uniformly over the whole projected sphere either, with no object farther than
2~pc behind the $l$=0~deg line; however, these stars distribute almost equally with nine ahead and eight
behind the Sun. Nevertheless, and until the deepest and most accurate searches have fully scrutinized the
6.5~pc-radius sphere (or larger spheres), we suggest that the number of brown dwarfs may be incomplete.

(ii) All-sky surveys such as 2MASS and WISE are complete over most of the sky. However, in the densest
regions towards the Galactic centre and plane and in globular clusters, there is confusion noise from too
many sources, which reduce the survey depth limit, for example by about 1.5--2~mag for
2MASS\footnote{\url{www.astro.caltech.edu/~jmc/2mass/v3/gp/analysis.html}}; besides this, bright stars with
saturated counts, diffraction spikes, and halos complicate or impede the automatic selection and visual
verification of candidates. However, an asymmetry with respect to the distribution of the problematic
crowded regions in the two hemispheres separated by the blue dividing line in Fig.~\ref{skymap} is not
expected.

(iii) While searching for nearby late-type dwarfs by photometry and sky motions using WISE and 2MASS,
\citetads{2014ApJ...783..122K} found objects missed by \citetads{2014ApJ...781....4L}, and conversely. The
most notable example is their second highest motion object, the $J$=9.0~mag M7 dwarf
\object{WISE~J1540-5101} at 4.4~pc \citepads{2014ApJ...783..122K,2014A&A...567A...6P}. Therefore, even
while recovering all previously known nearby dwarfs, independent searches find different new candidates.
\citetads{2014ApJ...783..122K} point out that they were able to measure motions with enough accuracy only
for the fastest among the nearest TY-type dwarfs, because the apparent magnitudes of the dwarfs are so
faint that their astrometric position errors at different epochs are very large. The more recent search by
\citetads{2016ApJ...817..112S} gather a sample of new objects of not-so-small motions and about two
magnitudes fainter ($W2$$\sim$14~mag) than their previous sample \citepads{2014ApJ...783..122K}, however
the late-type candidates are farther than 8~pc from the Sun. Using the $W2$-band survey limit of 14.5~mag,
they rule out other objects as faint as \object{WISE~J0855$-$0714} at $\la$2.9~pc or new earlier Y-type
dwarfs at $\la$9.5~pc, and with proper motions of 0.25--15~arcsec~yr$^{-1}$. Brighter, new T-type dwarfs at
short distances would then be ruled out even more.

\begin{figure}[h!] \resizebox{\hsize}{!}{\includegraphics[angle=-90,trim=48 76 25
50bp,clip]{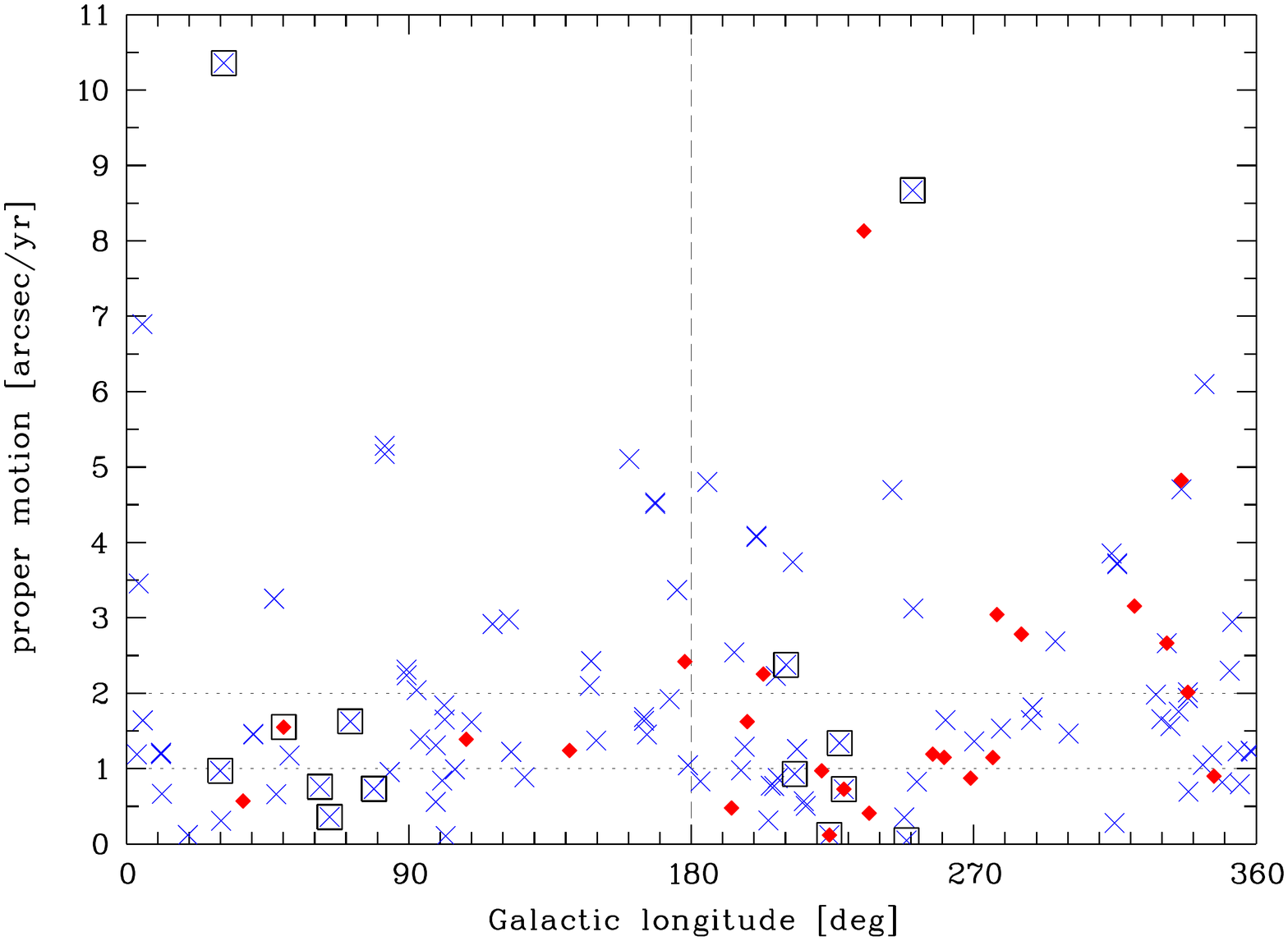}} \caption{Proper motions of 135 stars (crosses) and
26 brown dwarfs (filled lozenges) in the 6.5~pc sample as a function of Galactic longitude. The
dashed line marks the dividing line between the two hemispheres front and back of the Sun relative to
the direction of Galactic rotation. The brown dwarf distribution in the left part differs from that
in the right part, with a deficit at low ($\lesssim$1 arcsec/yr) and high ($\gtrsim$2 arcsec/yr)
proper motions indicated by dotted lines. Open squares indicate objects within 30~deg of the solar
apex- and antapex, respectively.} \label{pmlong} \end{figure}

Considering the proper motions of the stars and brown dwarfs in the 6.5~pc sample as a function of Galactic
longitude, we find a deficit of both small and large proper motions for the brown dwarfs with
$l\sim0-180$~deg (see Fig.~\ref{pmlong}). The median proper motion is always (for stars and brown dwarfs in
both hemispheres) at about 1.5 arcsec/yr. However, the mean and the standard deviation are about 2.0 and
1.8 arcsec/yr, respectively, except for the only five brown dwarfs at $l\sim0-180$~deg, for which these
values are reduced to 1.4 and 0.7~arcsec/yr, respectively. In addition to the deficit of low and very high
proper motion brown dwarfs at $l \sim 0-180$ deg, some brown dwarfs (and stars) with relatively small
proper motions near the solar apex- ($l=57, b=+22$~deg) and antapex directions ($l=237, b=-22$~deg) could
still be missing\footnote{This does not impede the two highest proper motion stars in the sky, Barnard's
and Kapteyn's stars, to be near the apex and antapex, respectively.}.


Thus, new brown dwarfs at $<$6.5~pc (especially at $l=0-180$~deg and $\la5$~pc) may still be discovered
eventually (a) at both small ($\lesssim$1 arcsec/yr) and large ($\gtrsim$2 arcsec/yr) proper motions or (b)
in the extremely crowded regions of the Galactic plane, or any sky location where blends or mismatches with
unrelated sources or discarded matches with related sources can occur. A near-infrared survey 3 to 4 mag
deeper than 2MASS such as the Visible and Infrared Survey Telescope for Astronomy (VISTA) -- south --
Hemisphere Survey and the UKIRT Infrared Deep Sky Survey (UKIDSS; see e.g. discoveries by
\citeads{2007MNRAS.381.1400W}; \citeads{2009MNRAS.395.1237B,2011MNRAS.414L..90B};
\citeads{2010A&A...515A..92S,2010A&A...510L...8S}; \citeads{2010MNRAS.405.1140G};
\citeads{2010MNRAS.408L..56L}; \citeads{2012MNRAS.422.1922P}; \citeads{2012A&A...548A..53L};
\citeads{2015ApJ...804...96G}), extended to the complete northern hemisphere, could be propitious.
Ultimately, some bright candidates might not be followed up spectroscopically yet, because of a lack of
timely observability. Figure~\ref{skymap} shows that the $l=0-90$~deg Galactic quadrant is accessible to
night observation during spring and summer, whereas the $l=90-180$~deg Galactic quadrant is accessible
during autumn and winter, from a northern observatory, such as at Mauna Kea.


(iv) If this observed non-uniform substellar distribution remains unchanged after future searches, then
either some fortuitous random aggregation or some dynamics affecting primarily the lower mass objects,
could be invoked. We note that pencil beam deep surveys of the Milky Way suggests there might be slightly
more M-type dwarfs in the Galactic northern hemisphere ($b>0$~deg) than in the southern hemisphere
($b<0$~deg; see \citeads{2014ApJ...788...77H} and references therein) and hints at a reversal occuring for
mid- to late M-type dwarfs, which would be more numerous in the southern hemisphere. However, as already
mentioned in Sect.\ref{results} and shown in Fig.~A.1, the objects in our sample distribute
uniformly perpendicular to the Galactic plane; indeed, we count 50 and 51 M dwarfs at $b\ge0$ and $<0$~deg,
respectively, and equal numbers of brown dwarfs. This is not surprising since the size of our sample (13~pc
diameter) is at least a few tens times smaller than typical disk scale heights. The cause of the planar
front/back asymmetry observed for the brown dwarfs is thus likely unrelated to that of the large scale
perpendicular asymmetry for stars, the latter asymmetry also derived from the Sloan Digital Sky Survey
(SDSS) or RAdial Velocity Experiment survey (RAVE) catalogues and attributed to wavelike gravitational
perturbations \citepads{2012ApJ...750L..41W,2013ApJ...777...91Y,2013MNRAS.436..101W}.


\section{Conclusions}


For the updated $<$6.5~pc sample, we find a star-to-brown-dwarf ratio of 136$/$26=5.2, while
\citetads{2012ApJ...753..156K} find a ratio of 6 (8~pc sample), and Henry et al. (2016) count 10 times more
stars than brown dwarfs\footnote{Abstract for AAS Meeting in Kissimmee, FL in January 2016, available at
RECONS web page.} (in their RECONS 10~pc sample). Because of the heliocentric and geocentric nature of the
distribution at $<$6.5~pc, the non-uniformity of the substellar distribution on the Galactic plane is
likely due to an observational bias (since we expect brown dwarfs to distribute uniformly as stars do). The
brown dwarf census would thus be incomplete. Assuming five more brown dwarfs ahead of the Sun, i.e. the
addition of two-thirds of the observed excess relative to uniformity, the star-to-brown-dwarf ratio would
decrease to 136$/$31=4.4$^{+1.0}_{-0.7}$, considering error bars corresponding to Poissonian uncertainties
of the number of brown dwarfs. It is slightly higher but agrees with the ratio
$N(0.08$$-$$1.0M_{\odot})/N(0.03$$-$$0.08M_{\odot}$)=3.3$^{+0.8}_{-0.7}$ estimated in the \object{Orion
Nebula cluster} \citepads{2004ApJ...610.1045S,2008ApJ...683L.183A}, noting also that the multiplicity
of the nearby stars may be better accounted for than in Orion (higher number of stars). As already pointed
out by \citetads{2012ApJ...753..156K}, the star-to-brown-dwarf ratii of the solar neighbourhood and star
forming regions may also differ because they consist of different dynamical environments with different
star formation histories and mass segregations. In the much less probable case the relative void of
substellar objects at $l$$\sim$0--180~deg (or excess at $l$$\sim$180--360~deg) is real, this could be
related to some random inhomogeneity or dynamical effect of trapping or deflection taking place for small
objects, so that they would be located more behind than ahead of the Sun relative to the direction of the
rotation of the Galaxy.

\begin{acknowledgements} 

We thank the referee for a constructive review that helped us to improve the paper. We thank the A\&A
Language Editor J. Adams for revising the English of the manuscript. This research made use of the VizieR
catalogue access tool and the SIMBAD database provided by the CDS Strasbourg, France, the M, L, T, and Y
dwarf compendium housed at \url{www.DwarfArchives.org}, and the solar neighbourhood census provided by
RECONS (\url{www.recons.org}).

\end{acknowledgements}

\bibliographystyle{aa}
\bibliography{/work2/gbihain/Publications/astro_ref/astronomy_ref_citeads}

\appendix

\section{Distribution of the nearest stars and brown dwarfs on the Galactic orthogonal plane}


\begin{center}
  \includegraphics[keepaspectratio=true,scale=1.0774]{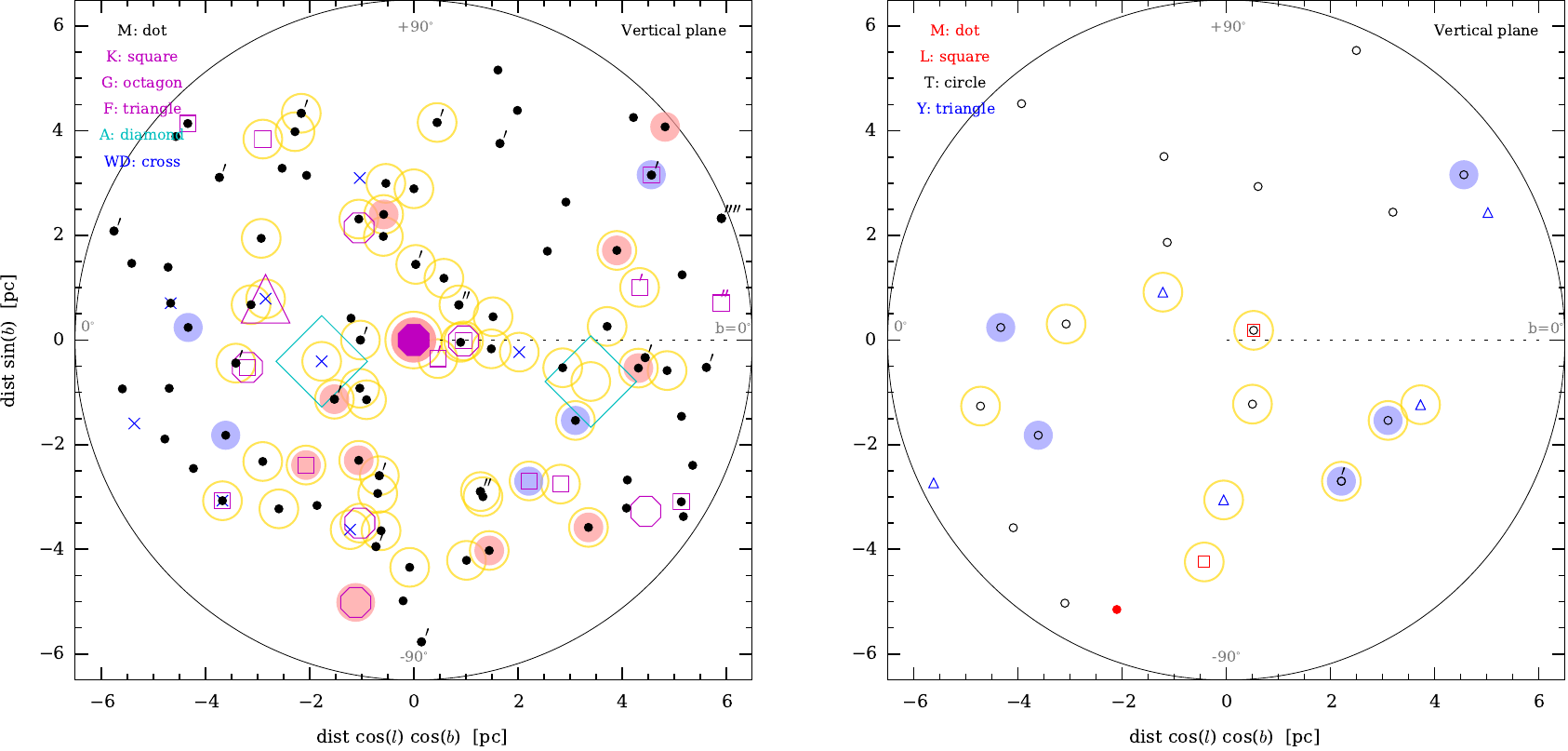}\label{distrz}
 \begin{minipage}{1.4\textwidth}
 \vspace{0.3cm}
 \noindent{\bf Fig. A.1.} Same as Fig.~\ref{distr}, but on the Galactic orthogonal plane along $l=0$~deg.
 \end{minipage}
\end{center}


\end{document}